\documentclass[conference]{IEEEtran}
\IEEEoverridecommandlockouts
\usepackage{cite}
\usepackage{amsmath,amssymb,amsfonts}
\usepackage{algorithmic}
\usepackage{graphicx}
\usepackage{textcomp}
\usepackage{xcolor}
\usepackage[flushleft]{threeparttable}
\usepackage{url}
\def\BibTeX{{\rm B\kern-.05em{\sc i\kern-.025em b}\kern-.08em
    T\kern-.1667em\lower.7ex\hbox{E}\kern-.125emX}}
\begin{document}

\title{A Dynamically Weighted Loss Function for Unsupervised Image Segmentation\\}

\author{\IEEEauthorblockN{1\textsuperscript{st} Boujemaa Guermazi}
\IEEEauthorblockA{\textit{Electrical, Computer, and Biomedical Engineering} \\
\textit{Toronto Metropolitan University}\\
Toronto, Ontario \\
bguermazi@ryerson.ca}
\and
\IEEEauthorblockN{2\textsuperscript{nd} Riadh Ksantini}
\IEEEauthorblockA{\textit{Computer Science} \\
\textit{University of Bahrain}\\
Zallaq, Bahrain \\
rksantini@uob.edu.bh}
\and
\IEEEauthorblockN{3\textsuperscript{rd} Naimul Khan}
\IEEEauthorblockA{\textit{Electrical, Computer, and Biomedical Engineering} \\
\textit{Toronto Metropolitan University}\\
Toronto, Ontario \\
n77khan@ryerson.ca}
\and

\thanks{This work was funded by the Natural Sciences and Engineering Research
Council of Canada.}
\thanks{Code for this project is available at: https://github.com/bijou-bijou/DynamicSeg}
}

\maketitle

\begin{abstract}
Image segmentation is the foundation of several computer vision tasks, where pixel-wise knowledge is a prerequisite for achieving the desired target. Deep learning has shown promising performance in supervised image segmentation. However, supervised segmentation algorithms require a massive amount of data annotated at a pixel level, thus limiting their applicability and scalability. Therefore, there is a need to invest in unsupervised learning for segmentation. This work presents an improved version of an unsupervised Convolutional Neural Network (CNN) based algorithm that uses a constant weight factor to balance between the segmentation criteria of feature similarity and spatial continuity, and it requires continuous manual adjustment of parameters depending on the degree of detail in the image and the dataset. In contrast, we propose a novel dynamic weighting scheme that leads to a flexible update of the parameters and an automatic tuning of the balancing weight between the two criteria above to bring out the details in the images in a genuinely unsupervised manner. We present quantitative and qualitative results on four datasets, which show that the proposed scheme outperforms the current unsupervised segmentation approaches without requiring manual adjustment.
\end{abstract}


\section{Introduction}
Image segmentation is fundamental to many application domains such as medical imaging, surveillance, self-driving cars, and sports. Image classification allocates a category label to the entire image. In contrast, image segmentation generates a category label for each input image pixel, dividing a whole picture into subgroups known as image segments. Although we have made notable progress, the segmentation process is still challenging due to various factors such as illumination variation, occlusion, and background clutters. 

Classical pioneering segmentation techniques such as active contour models (ACM) \cite{ACM}, k-means \cite{Kmeans}, and graph-based segmentation method (GS) \cite{graph} impose global and local data and geometry constraints on the masks. As a result, these techniques become sensitive to initialization and require heuristics such as point resampling, making them unsuitable for modern applications.

The current state-of-the-art models are deep learning methods based on convolutional neural networks (CNNs), especially the Mask R-CNN framework \cite{MRCNN}, which has been successfully put into practice for supervised semantic and instance image segmentation. Even so, such methods require a considerable amount of hand-labeled data, limiting their applicability in many areas. The problem becomes much more acute when it comes to pixel-wise classification, where the annotation cost per image is expensive. A possible solution to this problem is unsupervised image segmentation, where the image is automatically segmented into semantically similar regions. The task has been studied as a clustering problem in recent literature, reaching promising results. The Differentiable Feature Clustering \cite{them} is a state-of-the-art CNN-based clustering algorithm that simultaneously optimizes the pixel labels and feature representations \cite{them}. It applies a combination of \textit{feature similarity} and \textit{spatial continuity} constraints to backpropagate the model's parameters.  Feature similarity corresponds to the constraint that pixels in the same cluster should be similar to each other. Spatial continuity refers to the constraint that pixels in the same cluster should be next to each other (continuous).  However, to reach the desired segmentation result, \cite{them} applies a manual parameter tuning to find the optimal balancing weight $\mu$, which fails to achieve a good balance between the two aforementioned constraints depending on the degree of details in the image and dataset. This work introduces a novel dynamic weighting scheme that leads to a flexible update of the parameters and an automatic tuning of the balancing weight $\mu$. We achieve this by conditioning the value of $\mu$ to the number of predicted clusters and iteration number. We dynamically prioritize one of the constraints at each iteration to achieve a good balance. Experimental results on four benchmark datasets show that our method achieves better quantitative metrics and qualitative segmentation results by striking a better balance between feature similarity and spatial continuity. 

\section{Related Works}

\begin{figure*}[htb]

\hfill
\begin{minipage}[b]{1.0\linewidth} 
  \centering
  \centerline{\includegraphics[width=18cm]{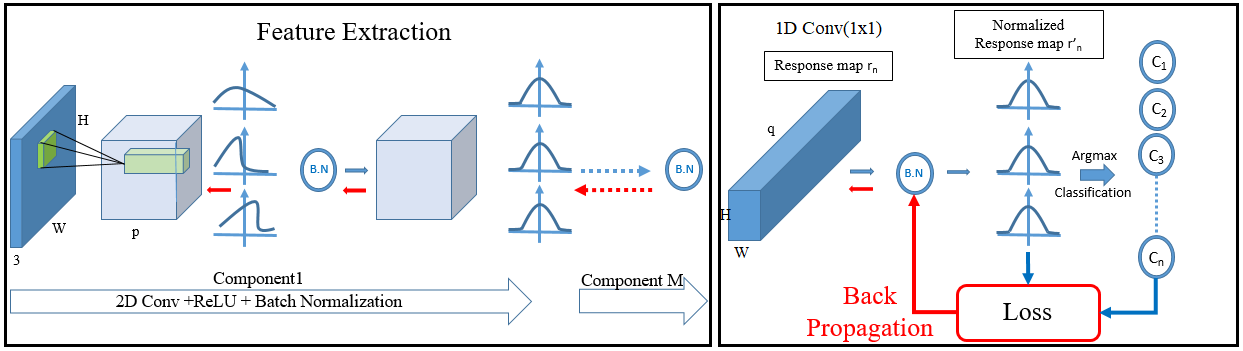}}
%  \vspace{1.5cm}
\end{minipage}
\caption{The CNN Framework: a forward-backward process which is iterated $T$ times to obtain the final prediction of the cluster labels $c_n$.}
\label{fig1}
\end{figure*}

Machine learning methods directly address the image segmentation problem by considering various features found in the image, such as colour or pixel information. K-means clustering \cite{Kmeans}, for instance, is a region-based method that divides an image into K groups based on the discontinuity properties of the extracted features. K-means is widely used for unsupervised segmentation. However, the hard clustering that assumes sharp boundaries between clusters in K-means does not guarantee continuous areas. The graph-based segmentation (GS) \cite{graph} is another region-based method that performs segmentation based on pixel similarity, distance, or colour weights. As a result, GS preserves details in low variability image regions while ignoring details in high variability regions, plus GS has a complex computation. On the other hand, The Invariant Information Clustering \cite{IIC} is an edge-based method and is easy to implement and rigorously grounded in information theory. The deep net IIC directly trains a randomly initialized neural network into semantic clusters without the need for postprocessing to cluster the high-dimensional representations.

Majority of the recent unsupervised segmentation models utilize deep learning. X. Xia et al. \cite{Wnet} tackles the problem of unsupervised image segmentation inspired by the concept of U-net architecture \cite{Unet}. They join two U-net structures into an auto-encoder followed by a post-processing phase to refine its prime segmentation.
The model merges segments using a Hierarchical Segmentation method to form the final image segments. It is a computationally expensive process and requires extensive hyperparameter tuning. L.  Zhou et al. proposed another neural network-based algorithm; the Deep Image Clustering (DIC) \cite{DIC}. DIC divided the image segmentation problem into two steps; A feature transformation sub-network (FTS) to extract the features first, then a trainable deep clustering sub-network (DCS) that groups the pixels to split the image into non-overlapping regions.
The DIC model has proven to be less sensitive to varying segmentation parameters and has lower computation costs. However, it uses superpixels to optimize the model parameters, which results in pre-determined fixed boundaries for segmentation regions.

In contrast to the approaches mentioned above that use intermediate representations followed by post-processing, W. Kim et al. propose utilizing a clustering algorithm that jointly optimizes the pixel labels and feature representations and updates their parameters using backpropagation \cite{them}. Furthermore, in contrast to the superpixel-based refinement process \cite{old}, the authors present a novel spatial continuity loss function to achieve dynamic segmentation boundaries as opposed to utilizing superpixels. It is a simple process that uses the backpropagation of the feature similarity loss and a new spatial continuity loss that resolves the problem caused by superpixels in \cite{old}. However, the weight coefficient for the loss function used must be adjusted each time according to the datasets and the degree of detail of the image. Instead, we propose a dynamic method for adjusting the weighting factor automatically, as described in the following sections.

\section{Proposed Method}

\subsection{The CNN framework}\label{AA}

For a fair comparison, we use the same network architecture and training framework used in \cite{them}. The model is shown in Figure (\ref{fig1}). $M$ convolutional components are used to produce a $p$-dimensional feature map $r$. The CNN subnetwork consists of $2D$ Conv layers, $Relu$ functions, batch normalizations, and a final linear layer classifier that classifies the features of each pixel into $q'$ classes. A batch normalization function is applied to the response map $r$ to get a normalized map $r'$. Lastly, the $argmax$ function is used to select the dimension that has the maximum value in $r'_n$. Each pixel is assigned the corresponding cluster label $c_n$, which is identical to allocating each pixel to the closest point among the $q'$ representative points. During the backward propagation, first, the loss $L$ (defined in the next section) is calculated. Then, the convolutional filters' parameters and the classifier's parameters are updated with stochastic gradient descent. This forward-backward process is iterated $T$ times to obtain the final prediction of the cluster labels $c_n$. The segmentation problem is handled in an unsupervised manner without knowing the exact number of clusters. The latter must be flexible according to the content of the image. Therefore, we must allocate a large number q to the initial cluster labels q'. After that, similar or spatially related pixels are iteratively integrated to update the number of clusters $q'$.

\subsection{Proposed dynamic weighting scheme}
The loss function in \cite{them} is designed to strike a balance between feature similarity and spatial continuity. %The forward process is to train the parameters of a fixed feature extraction function and a static mapping function in an unsupervised way and predict the unknown clusters ${c_n}$. The backward process is to update the parameters of both functions with a fixed ${c_n}$.
The loss function L from \cite{them} is shown in Equation (\ref{first equ}). 
\begin{equation}\label{first equ}
  L = L_{sim}(\{r'_n,c_n\}) + \mu L_{con}(\{r'_n\}),  
\end{equation}

where, $\mu$: Weight for balancing;  $L_{sim}$  : feature similarity;  $L_{con}$: spatial continuity; $c_n$: cluster labels ; $r'_n$: normalized response.

The loss function in equation (\ref{first equ}) consists of two parts. The first part is the feature similarity loss $L_{sim}$ which is the cross-entropy loss between the normalized response map $r'_n$  and the cluster labels ${c_n}$. Minimizing this loss would reveal network weights that ease the extraction of more accurate attributes for segmentation. Thus, $L_{sim}$ ensures that pixels of similar features should be assigned the same label. The second part is the Manhattan Distance $L1$ Norm of horizontal and vertical differences of the response map $r'_n$ as a spatial continuity loss which redresses the deficiency caused by superpixels \cite{old}. This additional loss component $L_{con}$ has proven to be efficient in removing an excessive number of labels due to their complex patterns or textures and ensuring that continuous pixels are assigned the same label.

\begin{figure}[htb]

\hfill
\begin{minipage}[b]{1.0\linewidth}
  \centering
  \centerline{\includegraphics[width=9cm]{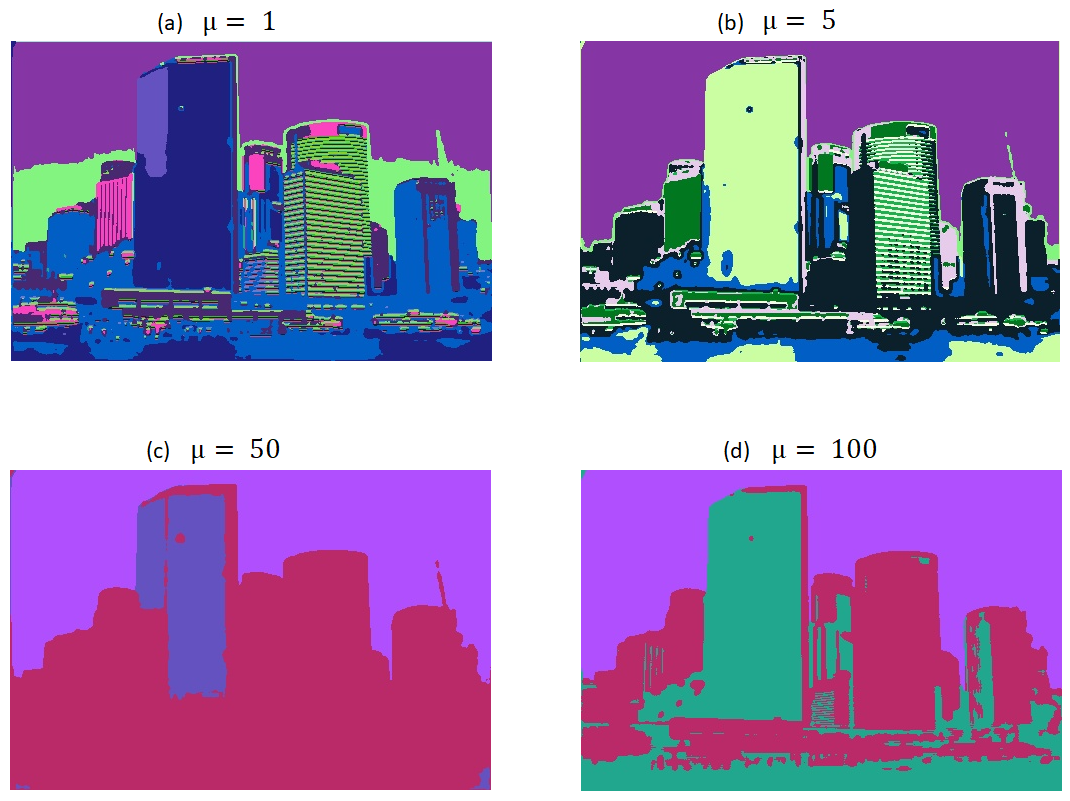}}
%  \vspace{1.5cm}
\end{minipage}
\caption{Results for different $\mu$ values on a sample image from the BSD500 dataset using the approach in \cite{them}.}
\label{fig2}
\end{figure}

While the loss L as mentioned above can result in reasonably accurate unsupervised segmentation results as reported in \cite{them}, the segmentation results are susceptible to the balancing parameter $\mu$. Figure (\ref{fig2}) show examples of the sensitivity to this parameter on an example image from the BSD500 dataset. As can be seen, for $\mu=50$ and $\mu=100$, the segmentation is coarse, resulting in sky, buildings, and coastal regions. However, the image is further segmented with $\mu= 1$ and $\mu= 5$, where buildings are further segmented into glass buildings, concrete buildings, and different floors. Although the authors argue that the value of $\mu$ is proportional to the coarseness of segmentation, We see that the results are not consistent, e.g. the segmentation for $\mu=50$ appears coarser than  $\mu=100$.

This poses a problem in practice. Such high sensitivity to the parameter means that for each dataset, this parameter has to be tuned extensively to obtain a result that is semantically more meaningful.

\begin{table*}[t]

\caption{Comparison of $mIOU$ for unsupervised segmentation on BSD500 and PASCAL VOC2012.
Best scores are in bold.}
\begin{center}
 \begin{threeparttable}
\begin{tabular}{|c|c|c|c|c|c|}
\hline
\textbf{}&\multicolumn{5}{|c|}{\textbf{dataset}} \\
\cline{2-6} 
\textbf{Method} & \textbf{\textit{BSD500 All}}& \textbf{\textit{BSD500 Fine}}&  \textbf{\textit{BSD500 Coarse}}&\textbf{\textit{BSD500 Mean}}&\textbf{\textit{PASCAL VOC2012}} \\
\hline
IIC \cite{IIC}  & 0.172 & 0.151 & 0.207 & 0.177 & 0.201 \\ 

\hline
k-means clustering & 0.240 & 0.221 & 0.265 & 0.242 & 0.238 \\ 

\hline
Graph-based Segmentation \cite{graph} & 0.313 & 0.295 & 0.325 & 0.311 & 0.286 \\ 

\hline
CNN-based + superpixels \cite{old}  & 0.226 & 0.169 & 0.324 & 0.240 & *** \\
\hline

%\cite{old} with superpixels & 0.226 & 0.169 & 0.324 & 0.256 & * \\
%\hline
 CNN-based + weighted loss, $\mu=5$ \cite{them} & 0.305 & 0.259 & 0.374 & 0.313 & 0.288\tnote{*}  \\ 

\hline
Spatial Continuity Focus $\mu'=100/q'$ & 0.329 & 0.288 & 0.406 & 0.341 & 0.289 \\ 

\hline
Spatial Continuity Focus $\mu'= 50/q'$ & 0.330 & 0.290 & 0.407 & 0.342 & {\bf0.290} \\ 
\hline
Feature Similarity Focus $\mu'= q'/10$ & 0.330 & 0.297 & 0.390 & 0.339 & 0.280 \\
\hline
Feature Similarity Focus $\mu'= q'/15$ & {\bf0.349} & {\bf0.307} & {\bf0.420} & {\bf0.359} & 0.275 \\
\hline

% copy& More table copy$^{\mathrm{a}}$& &  \\
% \hline
% \multicolumn{4}{l}{$^{\mathrm{a}}$Sample of a Table footnote.}
\end{tabular}

\begin{tablenotes}
       \item [*] All the results are copied from\cite{them}, except for the results on Pascal VOC 2012, as there is no indication in \cite{them} which images from the dataset were used for evaluation. Therefore, we chose 150 random images and re-produced the results for this dataset only. 
\end{tablenotes}
\end{threeparttable}

\label{tab1}
\end{center}
\end{table*}

In this work, we propose changing the weighting parameter's value during training dynamically. Our observation is that we can prioritize feature similarity during training at the earlier iterations and gradually shift focus to spatial continuity (or vice versa). We suggest a new dynamic loss function that includes a continuous variable $\mu$. The weight $\mu$ depends directly on the number of predicted clusters and iterations. We examine two versions of the proposed weighting scheme:

\begin{itemize}

\item Start by gathering continuous regions and shifting focus to feature similarity later. We call this approach \textit{Feature Similarity Focus (FSF)}. In this case, the performed trials lead to a linear function of the number of clusters (q') for the new dynamic balancing weight $\mu' = (q'/\mu)$ as shown in equation(\ref{equ3}). We tried other versions that vary exponentially with the value of $q'$; However, such functions resulted in a rapid change in the value of $\mu'$, which was not conducive to the balance we sought to achieve between the two constraints. 
\begin{equation}\label{equ3}
    L_{FSF} = L_{sim}(\{r'_n,c_n\}) + (q'/\mu)   L_{con}(\{r'_n\})
\end{equation}

\item Start by prioritizing feature similarity criteria early in training and end with a higher spatial continuity weight. We call this approach \textit{Spatial Continuity Focus (SCF)}. In this case, the proposed dynamic weight would be the multiplicative inverse of the number of clusters $\mu' = (\mu/q')$ as shown in equation (\ref{equ2}). Similar to FSF, we tried an exponential form, but the decay was too fast for it to be effective.
\begin{equation}\label{equ2}
    L_{SCF} = L_{sim}(\{r'_n,c_n\}) + (\mu/q')  L_{con}(\{r'_n\}) 
\end{equation}

\end{itemize}

\section{Experimental Results}

\begin{figure*}[!htbp]

%\hfill
%\begin{minipage}[b]{1.0\linewidth} 
  \centering
  \centerline{\includegraphics[width=17cm]{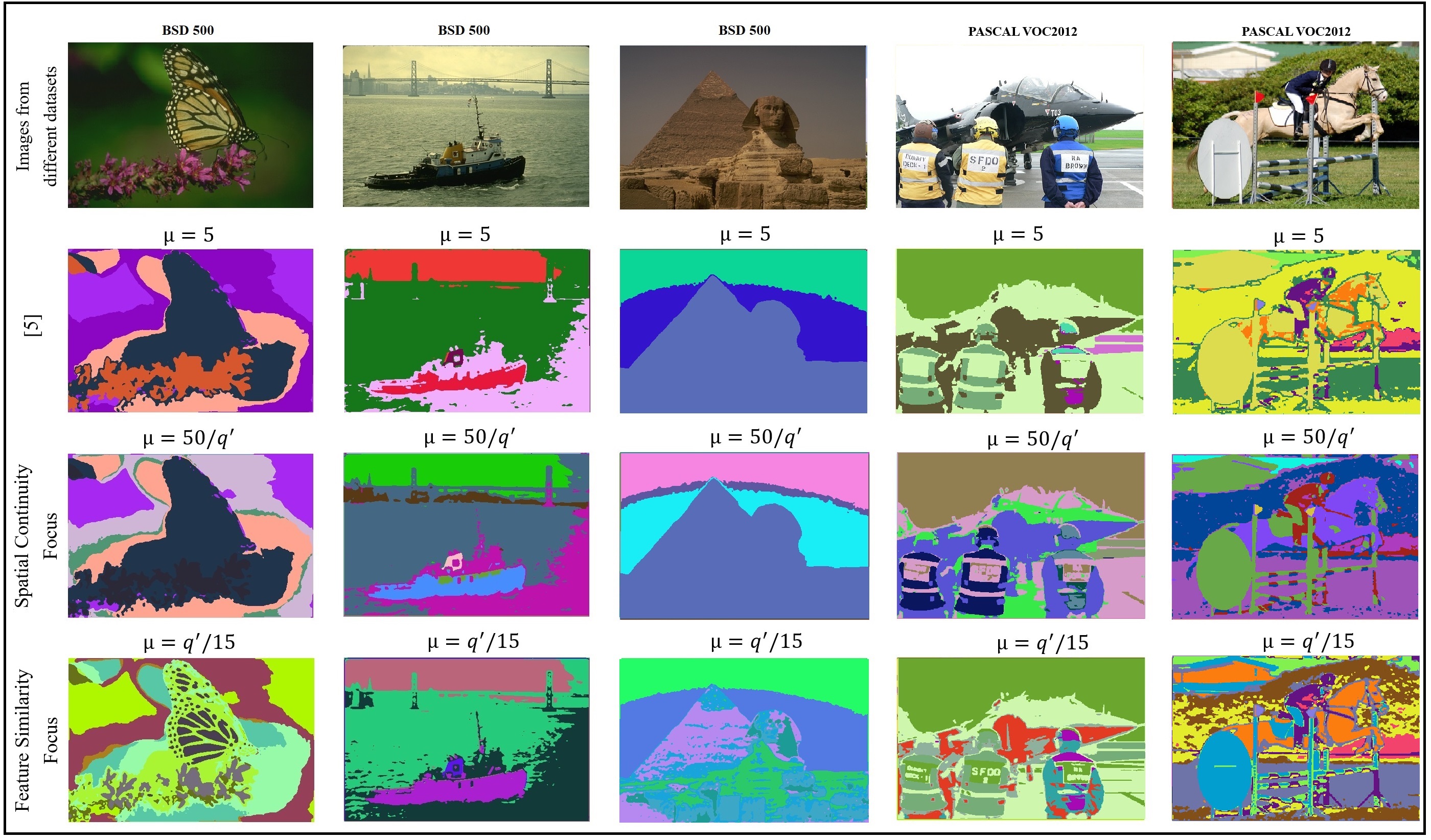}}
%  \vspace{1.5cm}
%\end{minipage}
%
\caption{Qualitative Results on select BSD500 and PASCAL VOC2012 images. Same color corresponds to the pixels being assigned the same clustering label by the algorithm. Please read Section \ref{qualres} for discussion on these results.}
\label{fig3}
\end{figure*}

\subsection{Experiment Setup}

The objective of the experiments is to show that our proposed dynamic weighting approach provides us with a more semantically meaningful segmentation. We replicate the experiments performed in \cite{them}. For all the tests, we fixed the number of components in the feature extraction phase $M$ to $3$. In addition, we set the dimension of the features space $p$ equal to the dimension of cluster space $q$ equal to $100$. Finally, we report the mean Intersection Over Union $mIOU$ overall images for the benchmark datasets. Ground truth is only used during the assessment phase and has no bearing on the training process.

Berkley Segmentation Dataset BSD500 \cite{BSD} and PASCAL Visual Object Classes 2012 \cite{pascal} are used to evaluate the segmentation results quantitatively and qualitatively. BSD500 consists of 500 color and grayscale natural images. Following the experimental setup in \cite{them}, we used the 200 color images of the BSD500 test-set to evaluate all the models. Since the BSD500 dataset contains multiple types of ground truth, we set three types of mIOU counting to assess the given results; "BSD500 All" takes into account all the ground truth files, "BSD500 Fine" considers the only ground truth file per image that has the most significant number of segments, and "BSD500 Coarse" takes only the ground truth file that contains the smallest number of segments. We defined "BSD500 Mean" as the average value of the above three measurements. For PASCAL VOC2012, we considered each segment an individual entity ignoring the object classification. VOC2012 is a large dataset containing 17,124 images with 2,913 images with semantic segmentation annotations. We randomly chose 150 of the semantic segmentation images to evaluate our method.

For the Icoseg \cite{Icoseg} and Pixabay \cite{pixab} datasets, select images are used to demonstrate additional qualitative results only. All of the aforementioned experimental settings are identical to the settings from \cite{them} for a fair comparison. 

\subsection{Quantitative results}

\begin{figure*}[htb]

\hfill
\begin{minipage}[b]{1.0\linewidth}
  \centering
  \centerline{\includegraphics[width=12cm]{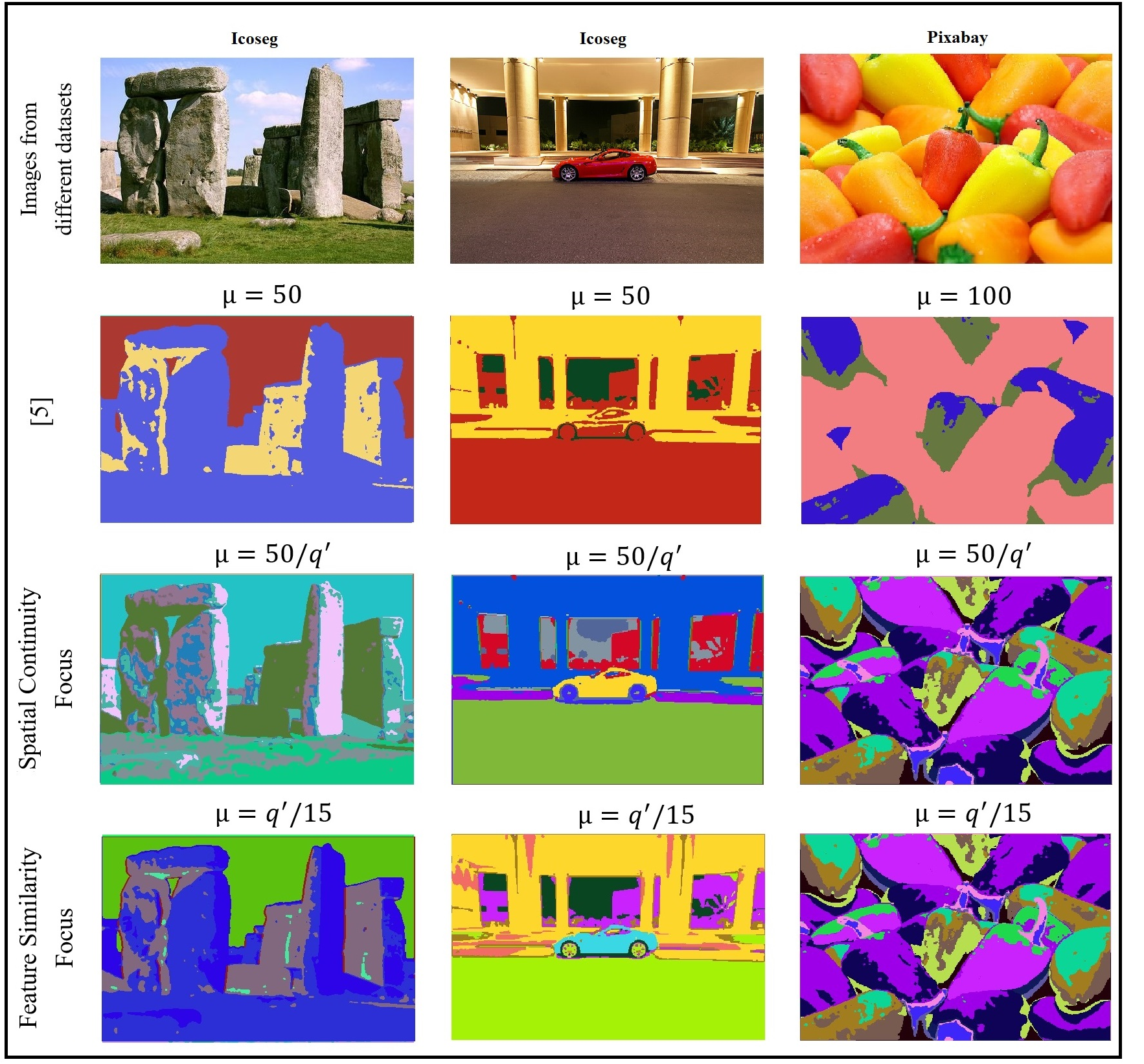}}
%  \vspace{1.5cm}
\end{minipage}
\caption{Qualitative Results on select Icoseg and Pixabay images. Same color corresponds to the pixels being assigned the same clustering label by the algorithm. Please read Section \ref{qualres} for discussion on these results.}
\label{fig4}
\end{figure*}

% We report the results  \cite{them}; $\mu=50$ for Icoseg \cite{Icoseg}, $\mu=5$ for BSD500 \cite{BSD}, and PASCAL VOC 2012 datasets \cite{pascal}, and $\mu =100$ for pixabay \cite{pixab}. In return for this, we used a continuous variable $\mu$ for each loss function in equation (\ref{equ2}) and equation (\ref{equ3}). First, we used equation (\ref{equ2}) to construct a model called similarity first, which starts by favoring feature similarity over spatial continuity and ends by reversing the power ratio in favor of continuity spatial. We tested two versions with $\mu=50/q'$ for all database, then $\mu=100/q'$ for all databases. Then we create the Continuity model using equation (\ref{equ3}). A model that proceeds by giving an important value to the spatial continuity, which will then be withered in favor of the similarity of the characteristics. Similarly, we tried two balancing weight values ($\mu=q'/10$ and $\mu=q'/15$) for all databases.
%Unlike the method proposed in \cite{them}, which trains the model per image with different initializations each time then chooses the best-obtained results, we have trained the model on 200 images from BSD 500 dataset \cite{BSD}, then on 150 images from PASCAL VOC2012 \cite{pascal}  with the same initial parameters. 

To demonstrate the effectiveness of the dynamic loss function, we compare the results with The Invariant Information Clustering (IIC) \cite{IIC}, the k-means clustering \cite{Kmeans} and the graph-based segmentation (GS) \cite{graph} in addition to the two CNN-based methods in \cite{old} and \cite{them}. One important note is that we report the same results as in \cite{them}, except for the results of the method \cite{them} on Pascal VOC 2012, as there is no clue about how the images were selected. In addition, the method is sensitive to the random initialization of the neural network weights, resulting in different mIOU scores every time we re-run the experiment. Therefore, we could not exactly reproduce the results reported in the paper using their provided code base. For a fair comparison, we fixed the initialization to a single set of values for all methods, including ours.

The Differentiable Feature Clustering \cite{them} uses a fixed value of $\mu$. Yet five is the ideal value of $\mu$ for BSD500 \cite{BSD}, and PASCAL VOC 2012 datasets \cite{pascal}, as that achieved the best results as reported in \cite{them}. In contrast, for our proposed SCF and FSF methods, the old $\mu$ is still the only hyper-parameter that needs to be tuned, but it is integrated into a dynamic loss function, as shown in equations (\ref{equ2}) and (\ref{equ3}). The best $\mu$ for SCF and FSF clustering were experimentally determined from \{25, 45, {\bf50}, 55, 60, 75 100, 200\} and \{2, 10, {\bf15}, 25,50, 100\}, respectively. We report results for $\mu$ = 10 and $\mu$ = 15 for FSF clustering ; $\mu$ = 100 and $\mu$ = 50 for SCF clustering. 

As illustrated in Table \ref{tab1}, the graph-based segmentation (GS) method outperformed the Differentiable Feature Clustering \cite{them} on “BSD500 all” and “BSD500 fine”. However, our proposed method outperforms \cite{old}, IIC, k-means, and, in particular, \cite{them} and GS for both datasets obtaining the best mIOU scores. While FSF shows the best performance for the BSD500 dataset, SCF achieves the best scores in the Pascal VOC2012 dataset and the second-best score in the BSD500 dataset.

\subsection{Qualitative results}
\label{qualres}
We also provide qualitative results on a few images as done in \cite{them}. As shown in Figure (\ref{fig3}), our model is more effective in bringing out segmentation regions that are semantically related. For example, For the "Show Jumping" image (column 5), the horse and the obstacle are classified as the same class by \cite{them} (both yellow). However, for both FSF and SCF, the horse and the obstacle are appropriately distinguished. For the "ship" image (column 2), \cite{them} fails to differentiate between the sky and the body of the ship (both red), but both proposed SCF and FSF can do it successfully.

Further qualitative results are shown for select Icoseg \cite{Icoseg} and Pixabay \cite{pixab} datasets that were also used in \cite{them}. These results can be seen in Figure (\ref{fig4}). The qualitative results on these datasets are presented to further demonstrate that our proposed approaches do not require as much parameter tuning as \cite{them} does. Figure (\ref{fig4}) highlights that the baseline Differentiable Feature Clustering \cite{them} is quite parameter sensitive. For each dataset, the weighting balance $\mu$ must be tuned extensively to obtain a more semantically meaningful result. For instance, PASCAL VOC 2012 and BSD500 datasets require a small balancing value $\mu=5$. While, Icoseg \cite{Icoseg} and Pixabay \cite{pixab} datasets need a much larger balance value $\mu=50$ and $\mu=100$, respectively. On the other hand, As illustrated in Figures (\ref{fig3}) and (\ref{fig4}), our proposed method has proven effective in dealing with different datasets using the same weight for both FSF and SCF. The proposed methods also bring out details in an unsupervised manner that is semantically more meaningful. For example, using the Feature Similarity Focus method (FSF), the red car from the iCoseg dataset (column 2 row 4 in Figure (\ref{fig4})) displays more detail on the tires and more precise building outlines than the details extracted by \cite{them} (column 2 row 2), where the car is partially blended into the road. Similarly, for the peppers image (column 4 in Figure (\ref{fig4})), \cite{them} was unable to identify the shapes of the individual peppers accurately. Both of our proposed method do a much better job, even with the same value of $\mu$ as the other images. 

Comparing the proposed SCF and FSF, SCF shows potency in class segmentation, while FSF performs better in object segmentation tasks. For instance, in the third row in Figure (\ref{fig3}), the SCF method segmented the image into the sky, dust, and monuments. However, the FSF segmented the image into the sky, dust, pyramid, and Sphinx. Depending on the application context, one of these approaches could be more favorable.

\section{Conclusion}

In this paper, we propose a dynamic weighting scheme for CNN-based unsupervised segmentation. Our proposed FSF and SCF approaches can strike a balance between feature similarity and spatial continuity for better segmentation. Qualitative and quantitative results on four datasets show that our proposed approach outperforms classical methods for unsupervised image segmentation such as k-means clustering and a graph-based segmentation method, and outperforms the state-of-the-art deep learning model, the Differentiable Feature Clustering, \cite{them} while producing semantically more meaningful results.

%In this paper, we propose a dynamic weighting scheme for CNN-based unsupervised segmentation. Our proposed FSF and SCF approaches can strike a balance between feature similarity and spatial continuity for better segmentation. Qualitative and quantitative results on two datasets show that our proposed approach outperforms a constant weighting scheme while producing semantically more meaningful results. 

\bibliographystyle{IEEEbib}
\bibliography{refs}

\end{document}